# Optical Fault Injection Attacks against Radiation-Hard Registers

Dmytro Petryk[1], Zoya Dyka[1], Roland Sorge[1], Jan Schäffner[1] and Peter Langendörfer[1,2]

[1]*IHP – Leibniz-Institut für innovative Mikroelektronik,* Frankfurt (Oder), Germany
[2]*BTU Cottbus-Senftenberg,* Cottbus, Germany
{petryk, dyka, sorge, schaeffner, langendoerfer}@ihp-microelectronics.com

*Abstract* — If devices are physically accessible optical fault injection attacks pose a great threat since the data processed as well as the operation flow can be manipulated. Successful physical attacks may lead not only to leakage of secret information such as cryptographic private keys, but can also cause economic damage especially if as a result of such a manipulation a critical infrastructure is successfully attacked. Laser based attacks exploit the sensitivity of CMOS technologies to electromagnetic radiation in the visible or the infrared spectrum. It can be expected that radiation-hard designs, specially crafted for space applications, are more robust not only against high-energy particles and short electromagnetic waves but also against optical fault injection attacks. In this work we investigated the sensitivity of radiation-hard JICG shift registers to optical fault injection attacks. In our experiments, we were able to repeatable trigger bit-set and bit-reset operations changing the data stored in single JICG flip-flops despite their high-radiation fault tolerance.

## I. Introduction

Nowadays tamper resistance is paramount for devices that collect, process and transmit private data, for example Internet of Things (IoT) devices. Data integrity and confidentiality have to be guaranteed. Additionally, such devices have to be resistant against different side channel analysis (SCA) attacks as well as malicious manipulations, e.g. fault injection (FI) attacks. The goal of FI attacks is to inject an error, which may modify stored data and/or turn the device into an unintended operation mode.

In laser-based attacks an adversary exploits the sensitivity of semi-conductors to light. These attacks belong to the class of semi-invasive attacks and are known as optical FI attacks [1]. In [2] we reported about successful optical FI attacks against different chips based on standard library cells manufactured in IHP's 250 nm and in IHP's 130 nm technology using the Riscure multi-mode red light laser source [8].

Devices with improved reliability seems to be more resistant against various disturbances. For example, radiation hard designs developed for space environment or for high-energy physics experiments are more robust against the natural disturbances caused by high-energy particles. Various radiation-hard techniques have been developed and verified, e.g. hardware Triple Modular Redundancy (TMR) [3], Dual Interlocked Storage Cell (DICE) [4], Junction Isolated Common Gate (JICG) [5], etc. These radiation-hard techniques show significant improvement of fault tolerance to radiation, i.e. higher level of reliability, in comparison to standard – non-radiation-hard – designs.

The ability of the radiation-hard designs to withstand the influence of short electromagnetic waves and high-energy particles suspects that such designs can be more robust against any electromagnetic disturbances including laser FI attacks. In this work, we verified the resistance of radiation-hard JICG shift registers to front-side optical FI attacks.

The paper is structured as follows. Section II describes the JICG radiation-hard technique. Section III describes the attacked JICG shift registers. Section IV describes the FI setup used. Section V discusses the results of the optical FI attacks performed. Section VI concludes this paper.

## II. JICG Radiation-Hard Technique

JICG is a radiation hardening by design technique. It is intended to be used in high energy physics experiments and in harsh environment such as outer space. JICG is designed to prevent single event upsets (SEUs), i.e. a change of the logical state of a flip-flop (FF) caused by high-energy particles, and to improve total ionizing dose (TID), i.e. collection of a charge in oxide due to radiation. Electric circuits of standard and JICG inverter gates are shown in **Fig. 1**–*(a)* and **Fig. 1**–*(b)*, respectively. **Fig. 1**–*(c)* shows the layout of a JICG inverter gate in IHP's 250 nm technology.

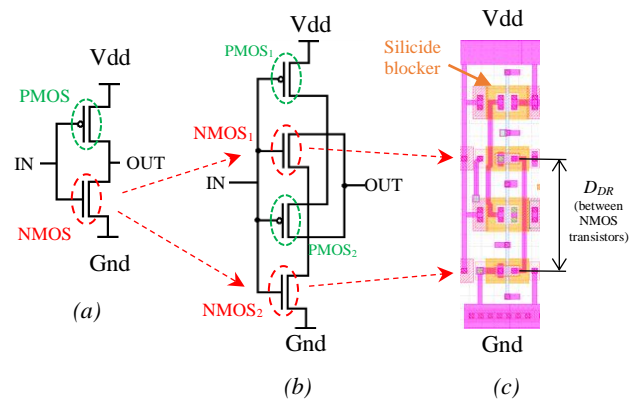

**Fig. 1.** Inverter gate:
*(a)* – electric circuit of standard inverter gate;
*(b)* – electric circuit of JICG inverter gate;
*(c)* – layout of JICG inverter gate.

To improve the TID the junction isolation (JI) of the drain-source region is realized using silicide blockers [5]. This allows to suppress the leakage of current between drain-source regions. To prevent SEUs redundancy of transistors is implemented. In particular, each transistor in a CMOS circuit is duplicated, i.e. each transistor is substituted by two transistors:

- instead of the single NMOS transistor in **Fig. 1**–*(a)*, a pair of transistors $NMOS_1$ and $NMOS_2$ is used, see **Fig. 1**–*(b)*;

- instead of the single PMOS transistor in **Fig. 1**–*(a)*, a pair of transistors PMOS$_1$ and PMOS$_2$ is used, see **Fig. 1**–*(b)*.

The gates of each duplicated transistor pair are connected, i.e. the duplicated transistors have a common gate (CG) and share the same conduction type. Such transistor redundancy allows to block the influence of a high-energy particle in any NMOS/PMOS branch of a CMOS circuit. To ensure the blocking capability a sufficient distance between the drain regions $D_{DR}$ of MOS transistors in a branch has to be kept. The distance $D_{DR}$ between the drain regions of the duplicated NMOS transistors is equal to the one of the duplicated PMOS transistors. In the IHP 250 nm technology this distance is $D_{DR} = 9$ µm. This sufficiently large $D_{DR}$ allows to keep one of the duplicated transistors unaffected if a high-energy particle hits the gate. For example, the transistor NMOS$_1$ works properly if a high-energy particle causes a fault in the transistor NMOS$_2$.

The realization of JICG MOS transistors corresponding to [5], [6] shows a significant improvement of radiation tolerance. However, the redundancy of transistors:
- requires additional area on a die that increases production costs;
- increases power consumption that can be an issue for battery powered devices;
- reduces switching speed [5], i.e. decreases the performance.

### III. ATTACKED CHIP

To assess the sensitivity of our JICG design to optical FI attacks we experimented with shift registers. The attacked shift registers are placed in Ceramic Dual In-line (C-DIP) package with a window, i.e. the front-side surface of the attacked chip is accessible and a decapsulation is not required. **Fig. 2** shows the attacked JICG chips in their C-DIP package.

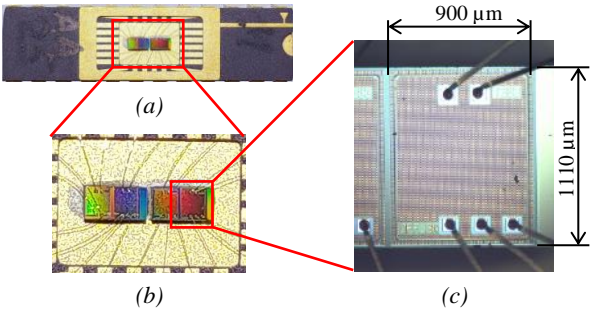

**Fig. 2.** The attacked chip:
*(a)* – 4 JICG shift registers in a C-DIP package;
*(b)* – 4 JICG shift registers, zoomed in;
*(c)* – a single JICG shift register, zoomed in.

The attacked chips contains 4 integrated circuits. Each of 4 integrated circuits is a 256 bit JICG register, i.e. each integrated circuit consists of 256 JICG flip-flops connected in series. A JICG flip-flop consists of 4 two-input and 2 three-input NAND gates, i.e. each JICG flip-flop consists of 6 NAND gates in total. **Fig. 3** presents structural scheme *(a)* and layout *(b)* of a single JICG flip-flop in the attacked shift registers.

Due to the increased number of transistors, the size of a single JICG flip-flop is relatively big (82×20 µm$^2$) compared to a non-radiation-hard flip-flop (20×7 µm$^2$) available in the standard IHP 250 nm gate library [2].

Each NAND gate in a JICG flip-flop is based on JICG inverter gates. **Fig. 4** shows the electric circuit *(a)* and the layout *(b)* of a two-input NAND gate based on two JICG inverters.

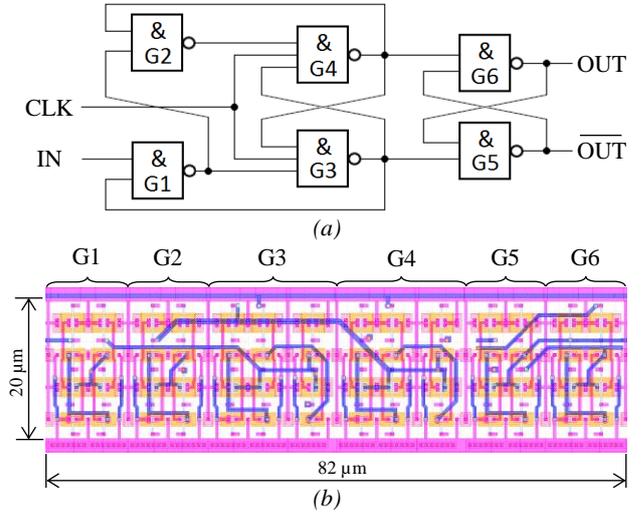

**Fig. 3.** JICG flip-flop in attacked shift registers:
*(a)* – structural scheme; *(b)* – layout view.

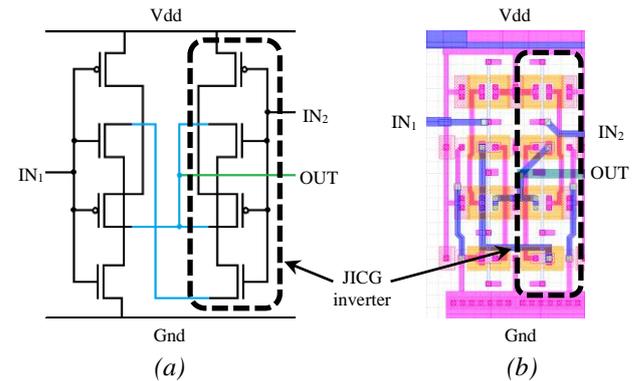

**Fig. 4.** A two-input NAND gate based on two JICG inverters:
*(a)* – electric circuit; *(b)* – layout view.

Each two-input NAND gate consists of 2 JICG inverters, i.e. it consists of 4 PMOS and 4 NMOS transistors. Each three-input NAND gate consists of 3 JICG inverters, i.e. it consists of 6 PMOS and 6 NMOS transistors.

The JICG shift registers attacked are manufactured in IHP's CMOS 250 nm radiation-hard technology (SGB25RH). Details of SGB25RH can be found in [6]. Chips manufactured in IHP's 250 nm technologies have 5 metal layers. Due to technology requirements, metal layers have metal fillers. Metal fillers are small metal structures that are placed in different metal layers to ensure mechanical stiffness during the manufacturing process. **Fig. 5** shows the front view of the attacked chip through the microscope using a 100× magnification objective. The image in **Fig. 5**–*(a)* is captured when focusing on Top Metal 2 and the image in **Fig. 5**–*(b)* is captured when focusing on Metal 3. Due to the metal fillers, the internal structure of attacked JICG shift registers is not visible through a microscope from the front-side. Metal fillers are obstacles for a visual inspection of the chip as well as for the laser beam, i.e. the metal fillers can significantly decrease the success rate of optical FI attacks [2].

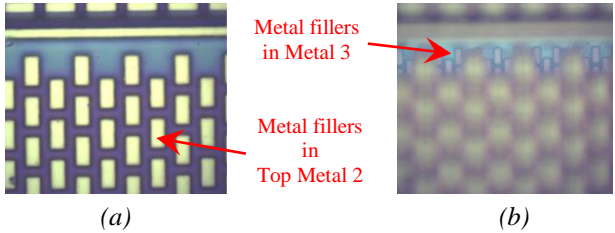

**Fig. 5.** Front view of the attacked chip using 100× magnification objective:
*(a)* – metal fillers in the top metal layer;
*(b)* – metal fillers in a middle metal layer.

In our experiments we expect FI to be successful due to the following facts:
- the diameter of the laser beam spot for the single-mode red laser can be set from 1 µm up to 15 µm using different magnification objectives (see details in section IV);
- the distance between metal fillers in the middle metal layer (see **Fig. 5**–*(b)*) is about 1.2 µm, i.e. the laser beam can reach the transistor level between metal fillers;
- the diameter of the laser beam spot using 5× magnification objective is 15 µm that is comparable to the distance $D_{DR} = 9$ µm, i.e. it is possible to influence either a PMOS transistor pair or an NMOS transistor pair but not both pairs simultaneously, see **Fig. 6**.

**Fig. 6** shows *(a)* laser beam spot size using a 5× magnification objective in comparison to the $D_{DR}$ on NAND gate layout and *(b)* electrical circuit of NAND gate with "closed" NMOS transistors that can be "opened" by the laser beam using the 5× magnification objective (see section IV).

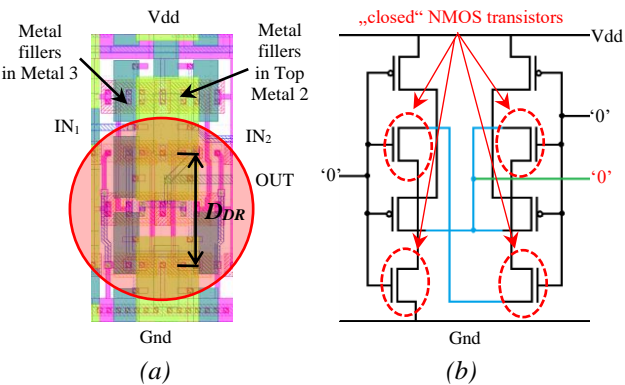

**Fig. 6.** NAND gate: *(a)* – schematic representation of laser beam spot size in comparison to the $D_{DR}$ on the layout using a 5× magnification objective; *(b)* – electrical circuit of a NAND gate with "closed" NMOS transistors that can be "opened" by the laser beam using 5× magnification objective.

Laser based attacks are feasible due to the internal photoelectric effect. This effect is observed in material with absent or low number of conduction electrons, i.e. dielectrics and semi-conductors. The internal photoelectric effect is based on the creation of conduction electrons in a silicon volume: when a laser illuminates the silicon electron-hole pairs are created along the beam. Under an electric field, the flow of these electrons creates a measurable increase in the current. So, using a laser beam it is possible to create conduction electrons and switch the transistor from the "closed" to the "open" state.

Since NMOS transistors are more sensitive to laser irradiation than PMOS transistors we expected that after successful FIs the number of "open" duplicated NMOS transistors will be higher than the number of "open" duplicated PMOS transistors. For example, a JICG NAND gate with closed NMOS transistor pairs has '0' as its logical inputs, and '1' as its output. By illuminating closed duplicated NMOS transistors with a laser beam using 5× magnification objective, it is possible to change the output from '1' to '0', i.e. the successfully attacked NAND gate in this case has the input as well as the output values '0', see **Fig. 6**–*(b)*. This switching of the NAND gate can cause a change of the value stored in the attacked flip-flop, i.e. the value stored in the attacked register will be manipulated.

We expected to observe in our experiments the following transient faults:
- *bit-set*: logical state '0' → logical state '1';
- *bit-reset*: logical state '1' → logical state '0';
- *stuck-at*: change of the cell state is no more possible till a reset of the device.

## IV. OUR EXPERIMENTAL SETUP

In our experiments we used Riscure equipment for laser FI [8], [10]:
- a Diode Laser Station (DLS);
- the Riscure Inspector FI software.

Additionally, we used a PC, a stable power supply, a signal generator and an oscilloscope. **Fig. 7** shows the structural view of the optical FI setup available at IHP.

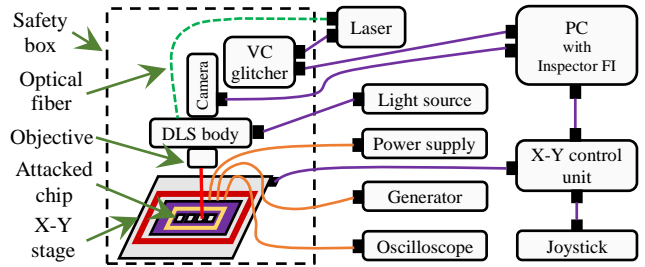

**Fig. 7.** Our experimental fault injection setup, structural view.

We used a modified 1st generation Riscure Diode Laser Station (DLS) [8]. In the modified system, mirrors inside the DLS body were replaced by the mirrors from 2nd generation of the Riscure DLS. These changes increase the transmitted laser beam power at the output of a magnification objective. The DLS consists of: a laser source, a control unit (VC glitcher), a source of light for chip illumination, a microscope camera, a DLS body, magnification objectives and an X-Y positioning stage.

Currently, one of three laser sources can be used for FI attacks in IHP: a multi-mode red laser with 808 nm wavelength, or a multi-mode infrared laser with 1064 nm wavelength, or a single-mode red laser with 808 nm wavelength (Alphanov Pulse on Demand Module (PDM) 2+ High Power (HP)) [7]. The Alphanov PDM has two laser sources with red (808 nm) wavelength. To increase the total laser beam output power the beams from both sources are propagated through the single-mode fiber. According to [7] and [9] the single-mode laser source has the following specifications:

- red wavelength with maximum power of 0.848 W;
- pulse duration in a range of 2 ns – continuous wave;
- circular spot sizes with a diameter $d$ of:
  - 15 µm (using 5× magnification objective);
  - 4 µm (20×);
  - 1.5 µm (50×);
  - 1 µm (100×).

Additionally, an X-Y positioning table manufactured by Märzhäuser Wetzlar GmbH & Co. [11] is a part of our set-up.

**Fig. 8** shows the modified Diode Laser Station with a single-mode Alphanov laser source used in our experiments.

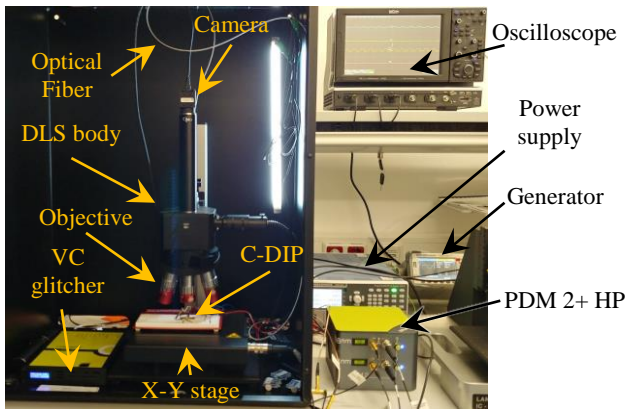

**Fig. 8.** Optical Fault Injection setup available at IHP.

The single-mode laser is connected to the DLS body through the proprietor coupling, i.e. laser source → single-mode optical fiber → coupling → DLS body. Since all the connections between PDM 2+ HP and the proprietor coupling are sealed, i.e. there is no leakage of laser light, the PDM 2+ HP is placed outside the safety box. The attacked chip with the JICG shift registers was placed on a breadboard on the Märzhäuser X-Y positioning table. The X-Y table moves the breadboard with the attacked chip relative to the laser beam.

## V. EXPERIMENTS AND RESULTS

In our experiments we use the single-mode laser source due to the following facts (compared to multi-mode lasers):

- Low laser beam output power can prevent damaging the attacked chips;
- Small laser beam spot size and known – here Gaussian – distribution of the intensity in the laser beam allow to illuminate selected inverter areas precisely and in case of successful FI to localize the sensitive gate(s)/area(s) more accurately.

### A. Initial conditions

To cover the area of the flip-flop evenly in our experiments we set the distance of 0.5 µm between two laser shots along the $x$ and $y$ axis. We use the Riscure Inspector FI software to define the coordinates of the first and the last shot points in the $x$ and $y$ line and the number of the laser shots within these distances along each axis. After the last shot in the current $x$ line the X-Y table moved to the first shot point in the next $x$ line. We denote the process when the X-Y table moves from the first to the last shot points in the $x$ and $y$ line as laser scanning. We denote laser scanning of a single JICG flip-flop with the same programmed laser beam parameters as an experiment. In our experiments we clocked the JICG shift registers with different frequencies: 2 MHz, 4 MHz, 7 MHz, 10 MHz and 20 MHz and we varied the laser pulse duration too. During an experiment we constantly send one logical state, e.g. only '0' to the input of shift register (or constantly only '1'), i.e. the input signal of the register was constant during an experiment.

### B. Determining successful FI attacks

To observe laser FI we used an oscilloscope connected to the output of the register. **Fig. 9** shows the oscilloscope waveforms of laser pulse, clock, input and output of the JICG register measured while attacking the last flip-flop, i.e. the flip-flop whose output is the output of the shift register. This allows observing each fault directly after its injection.

**Fig. 9**–*(a)* shows the case if no fault was injected, i.e. the input of the register is constantly '0' (see magenta line in **Fig. 9**–*(a)*) and the output signal is constantly '0' too (see yellow line in **Fig. 9**–*(a)*).

**Fig. 9**–*(b)* demonstrates a successful *bit-set* FI into the last flip-flop, i.e. the input of the register is constantly '0', but the output of the register is '1' during the clock cycle directly after the laser illumination.

**Fig. 9**–*(c)* demonstrates a successful *bit-reset* FI into the last flip-flop, i.e. the input of the register is constantly '1', but the output of the register is '0' directly after the laser shot.

### C. Experiment sequence

We selected randomly other single flip-flops for the experiments. We performed scanning of the selected flip-flops with different parameters of DLS, i.e. we:

- programmed laser beam output power and pulse duration;
- set an area for laser scanning, i.e. we defined the coordinates of the first and the last shot points in the $x$ and $y$ line and the number of the laser shots within these distances along each axis;
- selected magnification objective.

If the attack (during an experiment) failed we increased the laser beam power/pulse duration and repeated the attack. If a fault was not injected applying the maximum laser beam power and pulse duration we changed the magnification objective. Once we succeeded in attacking a flip-flop we repeated the attack with the same set of the parameters to confirm attack repeatability. If we got a positive result we selected another JICG flip-flop in the same chip and repeated the attack routine. Then, we performed similar attacks with the two other chips available in the package.

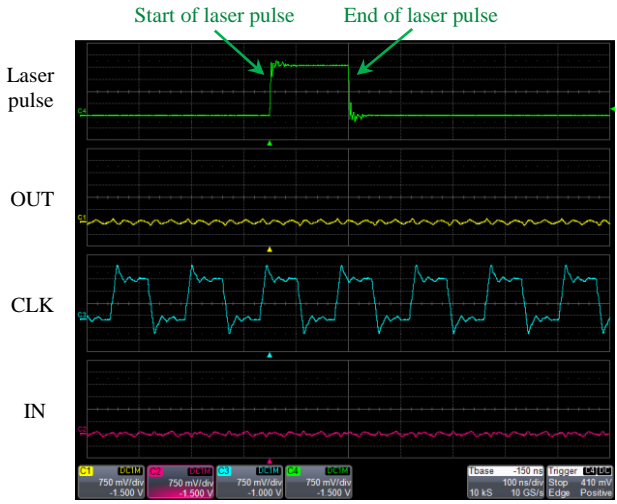

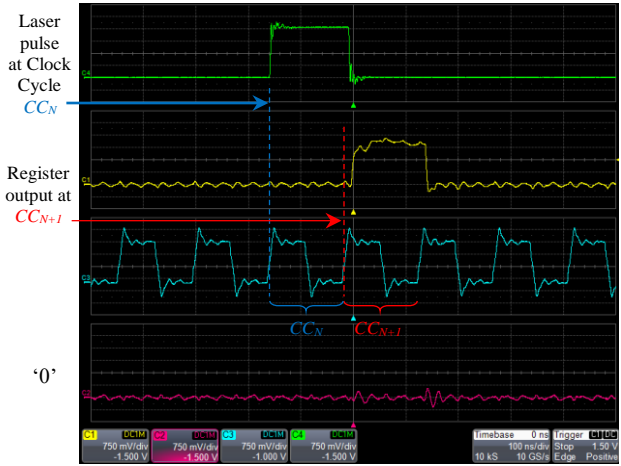

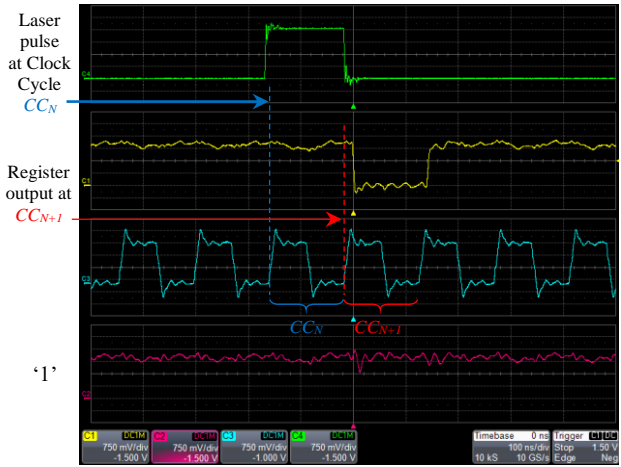

**Fig. 9.** Oscilloscope waveforms of laser pulse, clock, input and output of JICG register measured during scan of the last flip-flop: *(a)* – normal operation, i.e. FI attack failed; *(b)* – successful *bit-set*; *(c)* – successful *bit-reset*.

### D. Results of our FI attacks

According to the laser beam spot sizes given in [9] we should be able to influence a pair of duplicated NMOS/PMOS transistors in a single JICG inverter using 5× magnification objective only. Due to this fact and the unknown sensitivity of JICG registers to laser irradiation we began with a low laser beam power and short pulse duration using 5× magnification objective. We started with the following parameters: 10 % laser beam output power and 50 ns pulse duration.

In our experiments, all attacks using 5× magnification objective failed, even when attacking with the maximum laser beam power und pulse duration. We decided to perform attacks using 20×, 50× and 100× objectives to confirm inability to influence a pair of duplicated transistors simultaneously using other objectives, despite the fact that the laser beam spot size using these objectives is significantly smaller than $D_{DR}$. After an exhaustive scanning using all 3 objectives we observed successful FIs using the 20× magnification objective only. We successfully injected repeatable faults in flip-flops in the logical state '0' and the logical state '1', i.e. we injected *bit-set* and *bit-reset* faults using the 20× objective. We observed repeatable transient faults starting from 35 % laser beam output power and 50 ns pulse duration. Attacking other flip-flops, we observed successful FIs with a similar but not the same laser output power.

We performed additional experiments (i.e. laser scanning of a flip-flop) with increased laser beam power to assess how this may influence the behaviour of the JICG flip-flops. We observed only repeatable transient faults, i.e. neither permanent nor *stuck-at* faults were injected, even when attacking with maximum laser beam parameters. During our laser FI attacks we did not observe any damage of the chip surface.

### E. Results discussion

*1) Evaluation of the laser beam spot size*

According to [9] the diameter of the laser beam spot size using 20× magnification objective is $d = 4$ µm only, which is smaller than $D_{DR} = 9$ µm. For a successful FI it is necessary that the laser spot covers the gates of two duplicated transistors, i.e. it needs to be larger than $d \geq D_{DR} = 9$ µm, corresponding to the layout of the JICG flip-flop. Thus, we assumed that the laser beam spot sizes given in the datasheet [9] are not correct and we decided to evaluate this fact experimentally. Applying different magnification objectives we performed laser shots on the substrate surface and we measured the diameter of the reflections of laser beam spots. **Fig. 10** shows the reflections of the laser beam spots and their sizes we measured.

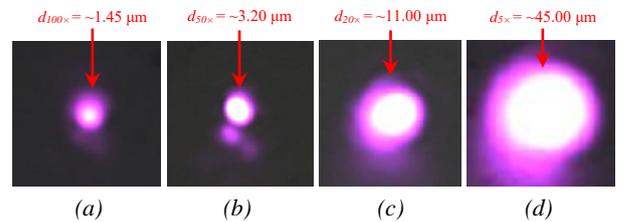

**Fig. 10.** Reflection of laser beam spots on the substrate surface using: *(a)* – 100× magnification objective; *(b)* – 50× magnification objective; *(c)* – 20× magnification objective; *(d)* – 5× magnification objective.

The laser beam was focused on the substrate surface, i.e. we captured the smallest beam waists using different magnification objectives in our setup. The sizes of the laser beam reflections are determined relative to the areas captured by the microscope camera [8] using the knowledge about the number of camera pixels per 1 µm² for each magnification objective used. We are aware that our measurements do not give very accurate results, especially if the expected laser spot size is about 1 µm using 100× objective. But the accuracy is reasonable for the evaluation and fits to our observations concerning the fault injections.

According to our observations, sizes of the reflected laser beam spots are significantly bigger than those given in [9]. **TABLE I** compares the laser beam spot sizes given in the Riscure datasheet [9] with the reflections of the laser beam spots observed in our experiments.

TABLE I. COMPARISON OF LASER BEAM SPOT SIZES GIVEN IN THE RISCURE DATASHEET WITH THE REFFLECTED SPOTS OBSERVED IN OUR EXPERIMENTS.

| Magnification objective | Laser beam spot size | |
|---|---|---|
| | Datasheet $d$, µm | Our observations[a] $\sim d$, µm |
| 100× | 1.00 | 1.45 |
| 50× | 1.50 | 3.20 |
| 20× | 4.00 | 11.00 |
| 5× | 15.00 | 45.00 |

[a.] Taking into account that values of the measured laser beam spot sizes given in the Riscure DLS datasheet [8] correspond to spot sizes where 80% of the energy is concentrated we expect that values given in [9] are measured in the same way.

The diameter of the laser beam spots with 100× and 50× objectives is significantly smaller than $D_{DR}$ = 9 µm. This can explain why our attacks with these objectives were not successful. The measured size of the laser spot with 20× objective explains also the success of our attacks: the diameter of the laser spot is 11 µm, i.e. it is large enough to cover the gates of two duplicated transistors. However, since the total laser beam output power is distributed across the laser beam spot, where its increase of in size is significantly higher than the increase of output power, lower intensity per 1 µm$^2$ is delivered. Thus, low intensity per square unit can be the reason of failed attacks using the 5× magnification objective.

*2) Details about the injected faults*

According to the JICG flip-flop layout, we were able to change the FF's logical state from '0' to '1', i.e. we induced *bit-set* fault, see **Fig. 9**–*(b)*, by illuminating gates G2, G5 (marked red in **Fig. 11**). Attacking the FFs in logical state '1' we were able to switch the FF's state to '0', i.e. we induced *bit-reset* fault, see **Fig. 9**–*(c),* by illuminating gates G1, G6 (marked yellow in **Fig. 11**). **Fig. 11** shows the sensitive areas of a JICG flip-flop that were revealed in our experiments. The areas marked red and yellow are sensitive to *bit-set* and *bit-reset* fault injections, respectively.

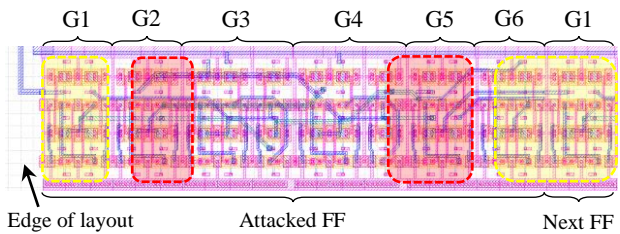

**Fig. 11.** Sensitive areas in attacked JICG flip-flop.

Please note that the accuracy of the sensitive areas determined is limited by the finite size of the laser beam spot, size of the metal fillers as well as inaccuracy in the X-Y table positioning, i.e. the position of the sensitive area can deviate up to 3 µm in each direction (the area shown in **Fig. 11** is 100×23 µm$^2$).

TABLE II summarizes the results of the optical FI attacks against JICG shift registers.

TABLE II. RESULTS OF ATTACKS AGAINST JICG SHIFT REGISTERS

| Register input | Power, %[a] | Pulse, ns | Magnification objective | Successful attack |
|---|---|---|---|---|
| '0' or '1' | ∞ | 50-10$^5$ | 100× | no |
| '0' or '1' | ∞ | | 50× | no |
| '0' | 35-100 | | 20× | bit-set |
| '1' | 45-100 | | | bit-reset |
| '0' or '1' | ∞ | | 5× | no |

[b.] Measurement unit of power in the Riscure FI Inspector software [10].

Using the 20× magnification objective the transient repeatable faults can be injected starting from 35 % up to 100 % of the laser beam output power and pulse durations from 50 ns up to 100 µm.

*F. Future work*

In our future work we plan to perform similar FI attacks using multi-mode laser sources with a higher beam output power than the one used in this work. We plan to place the metal fillers over sensitive areas of the flip-flops to evaluate them as countermeasure against laser FI attacks. For a successful FI the laser beam has to illuminate the gates of both duplicated transistors. Thus, if the gate of one of the duplicated transistors is covered by metal fillers it can prevent the laser FI from being successful. Since the placement of metal fillers is a mandatory step in design process covering the gates of selected transistors – using the metal fillers seems to be a reasonable low-cost countermeasure.

## VI. CONCLUSION

In this work we investigated radiation-hard JICG shift registers implemented in IHP's 250 nm technology to optical FI attacks. Despite their high level of reliability we were able to change the internal logical state of JICG flip-flops using a single-mode red laser source. We were able to inject *bit-set* as well as *bit-reset* faults. We determined the flip-flop's sensitive areas. In the future, the placement of metal fillers over gates of selected transistors has to be investigated as an effective countermeasure against laser FI attacks.

Since registers resistant to FI and SCA attacks are suitable for designing of cryptographic implementations the resistance of JICG registers against SCA attacks also has to be evaluated in the future. It is not excluded that JICG registers can be used as a basis for designing of logical gates resistant against a broad spectrum of physical attacks.

ACKNOWLEDGMENT

This project has received funding from the European Union's Horizon 2020 research and innovation program under the Marie Skłodowska-Curie grant agreement No 722325.